\documentclass[letterpaper]{article} 
\usepackage{aaai24}  
\usepackage{times}  
\usepackage{helvet}  
\usepackage{courier}  
\usepackage[hyphens]{url}  
\usepackage{graphicx} 
\usepackage{natbib}  
\usepackage{caption} 
\usepackage{algorithm}
\usepackage{algorithmic}

\usepackage{subcaption}
\usepackage{multirow}
\usepackage{booktabs}
\usepackage{xspace}
\usepackage{amssymb}
\usepackage{amsmath}
\usepackage{makecell}
\usepackage{newfloat}
\usepackage{listings}
\DeclareCaptionStyle{ruled}{labelfont=normalfont,labelsep=colon,strut=off} 
\floatstyle{ruled}
\newfloat{listing}{tb}{lst}{}
\floatname{listing}{Listing}
\newcommand{\Name}{\texttt{SAME}\xspace}
\newcommand{\Tref}[1]{Table~\ref{#1}}

\newcommand{\Fref}[1]{Figure~\ref{#1}}
\title{SAME: Sample Reconstruction against Model Extraction Attacks}
\author{
    Yi Xie\textsuperscript{\rm 1}\thanks{This work was done at NTU as a visiting student.}, Jie Zhang\textsuperscript{\rm 2}, Shiqian Zhao\textsuperscript{\rm 2}, Tianwei Zhang\textsuperscript{\rm 2}, Xiaofeng Chen\textsuperscript{\rm 1}\footnote{Corresponding author}\\
}
\affiliations{
    \textsuperscript{\rm 1}Xidian University, China\\
    \textsuperscript{\rm 2}Nanyang Technological University, Singapore\\

    xieyi@stu.xidian.edu.cn,
    \{jie\_zhang, shiqian.zhao, tianwei.zhang\}@ntu.edu.sg,
    xfchen@xidian.edu.cn
}

\usepackage{bibentry}

\begin{document}

\maketitle

\begin{abstract}
While deep learning models have shown significant performance across various domains, their deployment needs extensive resources and advanced computing infrastructure. As a solution, Machine Learning as a Service (MLaaS) has emerged, lowering the barriers for users to release or productize their deep learning models. However, previous studies have highlighted potential privacy and security concerns associated with MLaaS, and one primary threat is model extraction attacks. To address this, there are many defense solutions but they suffer from unrealistic assumptions and generalization issues, making them less practical for reliable protection. Driven by these limitations, we introduce a novel defense mechanism, \Name, based on the concept of sample reconstruction. This strategy imposes minimal prerequisites on the defender's capabilities, eliminating the need for auxiliary Out-of-Distribution (OOD) datasets, user query history, white-box model access, and additional intervention during model training. It is compatible with existing active defense methods. Our extensive experiments corroborate the superior efficacy of \Name over state-of-the-art solutions. Our code is available at https://github.com/xythink/SAME.
\end{abstract}

\section{Introduction}

Deep learning models have demonstrated superior performance in various domains. Yet, they often demand significant resources, including vast training data, advanced computational capabilities, and rigorous parameter optimization efforts. These requirements make deep learning models invaluable and expensive for adoption. Consequently, Machine Learning as a Service (MLaaS) has garnered significant interest, offering users a simplified and cost-efficient avenue to deploy sophisticated models.

Despite these advantages, a significant body of research has also revealed the privacy and security risks of models deployed with MLaaS \cite{tramer2016stealing, shokri2017membership, liu2021machine, yang2023protect, lou2021naspy}. Among these, model-extraction attacks \cite{yu2020cloudleak, pal2020activethief, zhao2023extracting, chen2021stealing, li2022extracting} represent a prominent threat, posing a direct risk to the intellectual property rights of the model owner.
The objective of model-extraction attacks is to locally recreate the target model at a minimal cost, leveraging limited queries to the openly deployed victim model. To achieve this goal, earlier works have proposed different strategies to generate the query samples for efficient model stealing, including surrogate sample \cite{pal2020activethief, orekondy2019knockoff}, adversarial sample \cite{yu2020cloudleak, papernot2017practical}, and synthetic sample \cite{barbalau2020black, truong2021data, kariyappa2021maze}. These attacks exhibit high effectiveness and efficiency across different threat environments, deep learning models, and tasks. 

\begin{figure}[tp]
    \centering
    \includegraphics[width=\linewidth]{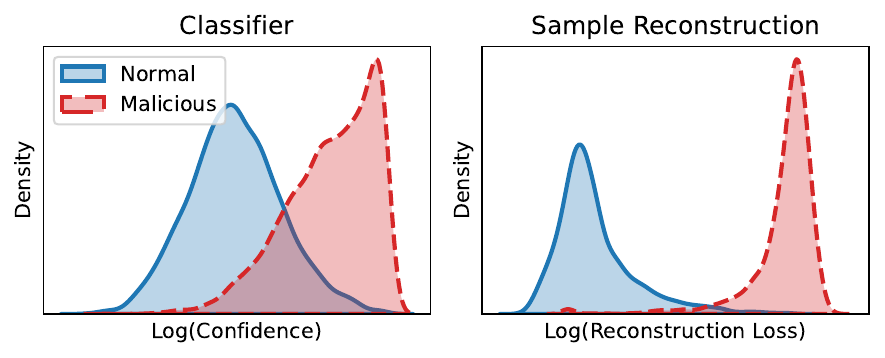}
    \caption{Distributions of anomaly scores for the classifier-based detection (left) and our sample reconstruction-based detection (right). The $x$-axis is in the logarithmic scale due to its long-tailed distribution. We utilize MNIST as normal query samples and employ KnockoffNets (with EMNIST-digits as the proxy set) to generate the malicious query samples. All samples undergo consistent preprocessing.
    }
    \label{fig:compare}
\end{figure}

Many efforts have been made to mitigate model extraction attacks \cite{jiang2023comprehensive}. 
Among them, malicious sample detection is the mainstream strategy. The model owner aims to distinguish the query samples used for model extraction from normal ones, and then reject them or return obfuscated responses. However, the advance and diversity of attack approaches pose several challenges in designing an effective detector. 
\textbf{First}, some defenses, such as Prada \cite{juuti2019prada}, require keeping a record of each user's queries for anomaly detection. They become vulnerable when the adversary launches a distributed attack \cite{yao2023fdinet}. 
\textbf{Second}, some approaches build machine learning classifiers to differentiate malicious and normal samples based on their features or predicted confidence scores \cite{kariyappa2020defending}. They are less effective in handling out-of-distribution (OOD) samples, even if the \textit{Outlier Exposure} (OE) strategy \cite{hendrycks2018deep} is adopted. As shown in Figure \ref{fig:compare} (left), the malicious samples have a large overlap with normal ones, indicating that lots of samples will be misclassified. 
\textbf{Third}, some methods leverage \textit{Ensemble of Diverse Models} (EDM) to detect malicious samples. They necessitate training several duplicates of the victim model using both in-distribution and auxiliary OOD samples, which significantly amplifies the defense costs. 

To address these challenges, we propose a novel detection method: \underline{S}ample reconstruction \underline{A}gainst \underline{M}odel \underline{E}xtraction (\Name). Our observation is that in-distribution and out-of-distribution samples exhibit significantly different features in the reconstruction process, as shown in Figure \ref{fig:compare} (right). This inspires us to leverage sample reconstruction to detect malicious queries. Specifically, \Name adopts the Masked Auto-encoder \cite{he2022masked} to reconstruct each query sample. It further builds an Auxiliary Model to repair the prediction of the reconstructed sample. Finally, it judges suspicious queries by combining two sources of information: reconstruction loss from the Masked Auto-encoder and deviation loss from the Auxiliary Model. 
Compared to existing works, \Name imposes minimal prerequisites on the defender's capabilities: (1) it eliminates the need for an auxiliary OOD training dataset; (2) it avoids retaining the user's query history; (3) it removes the demand for white-box access to the victim model. We conduct extensive experiments to demonstrate the superiority of \Name in detecting different types of model extraction attacks over SOTA methods. 

In summary, the main contribution of this paper includes three aspects: 
\begin{itemize}
    \item  We reveal the inherent weaknesses of classifier-based detection mechanisms, especially when confronted with unseen malicious queries in model extraction scenarios.
    \item We introduce \Name, a novel malicious query detection method rooted in sample reconstruction, significantly reducing the demands on defenders and acting as a versatile add-on to bolster current active defense strategies.
    \item  We demonstrate the effectiveness of \Name under multiple attack types through extensive experiments.
\end{itemize}

\section{Preliminaries}
In this section, we first introduce model extraction attack (MEA) and the corresponding detection methods. Then, the threat model is provided in detail.

\subsection{Model Extraction Attack}
Given a victim model $F_{V}$ (typically considered a black-box), the objective of a Model Extraction Attack (MEA) is to derive a functionally equivalent substitute model $F_{S}$ for illegal purposes (e.g., intellectual property violation). 
This could be formulated as minimizing the similarity loss on the victim model test set $D_V^{test}$:
\begin{equation}
    \underset{F_{S}}{\mathrm{min}} \sum_{x\in D_{V}^{test}} \mathcal{L}(F_{V}(x), F_{S}(x)),
\end{equation}
where $x$ denotes the samples and $\mathcal{L}$ is a loss function measuring the discrepancy between the outputs of $F_{V}$ and $F_{S}$.


The performance of the substitute model is highly affected by the query samples submitted by the attacker \cite{orekondy2019knockoff}. Past works have proposed different methodologies to construct query samples to improve extraction accuracy and efficiency. They can be classified into the following three categories. 

\noindent\textbf{Sampling-based Stealing.} This type of attack aims to construct a query dataset from a proxy dataset (often composed of public datasets) using a sampling strategy. Since different samples can provide different amounts of information to the substitute model, an appropriate sampling strategy can improve the attack performance. Knockoff \cite{orekondy2019knockoff} likens the sampling strategy to a multi-armed bandit problem in reinforcement learning. It adjusts the sampling strategy for the next step according to the reward from the previous actions. In addition, a series of works \cite{pal2020activethief, chandrasekaran2020exploring} use active learning to improve the stealing efficiency.

\noindent\textbf{Perturbation-based Stealing.} 
It was pointed out that samples lying approximately on the decision boundary of the victim model can greatly reduce the query cost \cite{yu2020cloudleak,wang2021delving}. Therefore some works introduce the perturbation-based strategy to generate query samples distributed near the decision boundary.
As a representative, JBDA \cite{papernot2017practical} proposes perturbation based on the Jacobian matrix on a small number of original victim training samples. CloudLeak \cite{yu2020cloudleak} uses a variety of adversarial perturbation methods to generate samples that approximate the model's decision boundary. Extensive experiments demonstrate the benefit of these perturbation strategies in boosting the stealing performance.

\noindent\textbf{Synthetic-based Stealing.} In most scenarios, the adversary does not have any dataset for model extraction. He can only generate noise samples for stealing \cite{truong2021data}. 
The mainstream strategy is to use the gradient approximation method \cite{truong2021data, kariyappa2021maze} to generate query samples, which can obtain more information regarding the victim model. 

\subsection{Model Extraction Attack Detection}

A popular defense direction is to detect the malicious query samples used for model extraction. This can be formulated as a binary classification problem. For each query, the model owner determines whether the sample is from OOD or not by calculating an anomaly score $S(x)$, and comparing it with 
a threshold $\lambda$. 
A higher $S(x)$ indicates a greater possibility that this sample $x$ is from $OOD$. It is important to minimize the misclassification of samples from $ID$ (normal queries).

There are different strategies for building such an anomaly detector. One possible direction is to detect the extraction activity at the \textbf{user level}. For instance, 
Prada \cite{juuti2019prada} keeps query logs for all users to spot potential suspicious activities. However, as pointed out by some works \cite{yao2023fdinet}, user-level detection based on the query history cannot mitigate distributed attacks, where the adversary employs multiple accounts to query the victim model. Hence, a more promising direction is \textbf{sample-level} defense \cite{kariyappa2020defending,kariyappa2020protecting,dziedzic2021increasing}, which performs detection on each sample. Existing solutions can be mainly divided into the following two types:


\noindent\textbf{Outlier Exposure (OE).} In practical settings, it is implausible for the model owner to have prior knowledge of the malicious query set $\mathcal{D}_{A}$. Therefore, some works~\cite{kariyappa2020defending} introduce an auxiliary outlier set $\mathcal{D}_{OE}$, which is disjoint from $\mathcal{D}_{A}$, to assist in learning a classifier for potential outliers. By exposing $\mathcal{D}_{OE}$ to $\emph{F}_{V}$ during training, OE makes $\emph{F}_{V}$ produce uniform probability distribution $\mathcal{U}$ on outliers. The optimization equation is:
\begin{equation}
    \mathbb{E}_{x\sim\mathcal{D} _{V}}\left[\mathcal{L} (\emph{F}_{V}(x),y)	\right]+\gamma\mathbb{E}_{x'\sim\mathcal{D} _{OE}}\left[\mathcal{L}_{OE} (\emph{F}_{V}(x'),\mathcal{U})\right],
\end{equation}
\noindent where $\mathcal{L}$ is the original learning objective, and $\mathcal{L}_{OE}$ is the outlier exposure loss.

However, OE needs to incorporate an auxiliary dataset into the training process of the victim model, which will introduce additional training overhead. Furthermore, additional learning objectives can degrade the accuracy of the model on the original task (see Table \ref{tab:classifier-acc}). 

\noindent\textbf{Ensemble of Diverse Models (EDM).} EDM utilizes an ensemble of diverse models $\left\{f_{i}\right\}_{i=1}^{i=N}$ to produce discontinuous predictions for OOD data~\cite{kariyappa2020protecting}. Similar to OE, EDM also leverages an auxiliary outlier set $\mathcal{D}_{OE}$ to defend against model stealing. Specifically, $\left\{f_{i}\right\}_{i=1}^{i=N}$ are trained jointly on $\mathcal{D}_{V}^{train}$ and $\mathcal{D}_{OE}$ according to the accuracy and diversity objectives:
\begin{equation}
    \mathcal{L} = \mathbb{E}_{x\sim\mathcal{D} _{V},x'\sim\mathcal{D} _{OE}}\left[(\frac{1}{N}\sum_{i=1}^{N}\mathcal{L}(\emph{f}_{i}(x),y)) + \gamma \mathcal{L}_{Div}(\emph{f}_{i}(x'))	\right],
\end{equation}
\noindent where the first loss term ensures the model utility on $\mathcal{D}_{V}$, and the second term ensures the diversity of predictions for a single outlier sample form $\mathcal{D}_{OE}$ across multiple submodels. Since this method has no explicit anomaly score, we compute the score based on the consensus among these diverse models, and it is smaller when models agree. This idea is also used in previous work \cite{dziedzic2021increasing}.

We shall point out that the effectiveness of detection based on auxiliary OOD datasets largely depends on the similarity between the distributions of auxiliary datasets and real malicious queries. The detector will perform poorly when such a distribution gap is large.





\subsection{Threat Model}
\noindent\textbf{Attacker's Ability and Goal.
}
As mentioned above, the goal of the attacker is to obtain his substitute model $F_S$, which is functionally similar to the victim model $F_V$. To achieve it, we assume that the attacker can leverage any MEA strategies. In this paper, we adopt three popular MEAs for implementation, namely, KnockoffNets (Knockoff) \cite{orekondy2019knockoff}, Jacobian-Based Dataset Augmentation (JBDA) \cite{papernot2017practical}, and Data-Free Model Extraction (DFME) \cite{truong2021data}.
Additionally, malicious queries submitted by attackers may vary in proportion to the overall queries (see \Tref{tab:ood-ratio}).

\noindent\textbf{Defender's Ability and Goal.}
In this paper, we mainly focus on defending MEA by detection methods. In other words, the defender shall detect the malicious query fed by the attacker.
Contrary to existing MEA detection methods \cite{kariyappa2020defending,kariyappa2020protecting}, 
we consider a more practical scenario as follows:
1) the defender does not need auxiliary OOD datasets or users' query history, and he cannot interfere with the victim model training process;
2) more importantly, the defender is unaware of the distribution of malicious queries under different attack strategies, namely, the defense is expected to be general over different types of model extraction. 




In a nutshell, the ability of the defender is more limited, inducing greater challenges for our MEA detection.

\begin{figure}[tp]
    \centering
    \includegraphics[width=\linewidth]{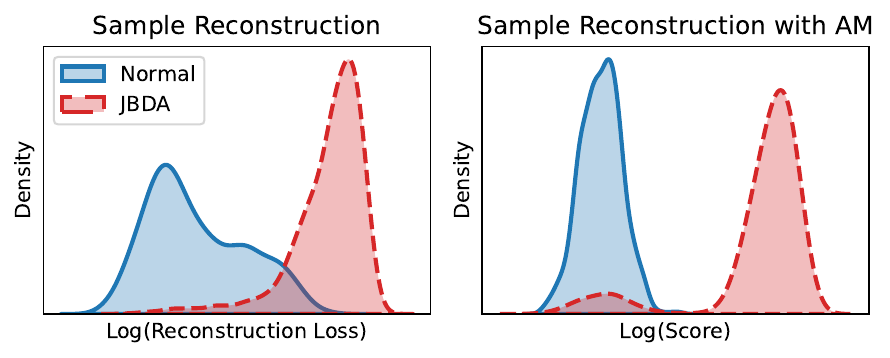}
    \caption{Distributions of anomaly scores for the reconstruction-based detection without (left) and with (right) Auxiliary Model (AM). The $x$-axis is in the logarithmic scale due to its long-tailed distribution. We utilize CIFAR-10 as normal query samples and employ JBDA (with 200 seed samples) to generate the malicious query samples. All samples undergo consistent preprocessing.
    }
    \label{fig:compare-jbda}
\end{figure}

\begin{figure*}[htp]
    \centering
    \includegraphics[width=\linewidth]{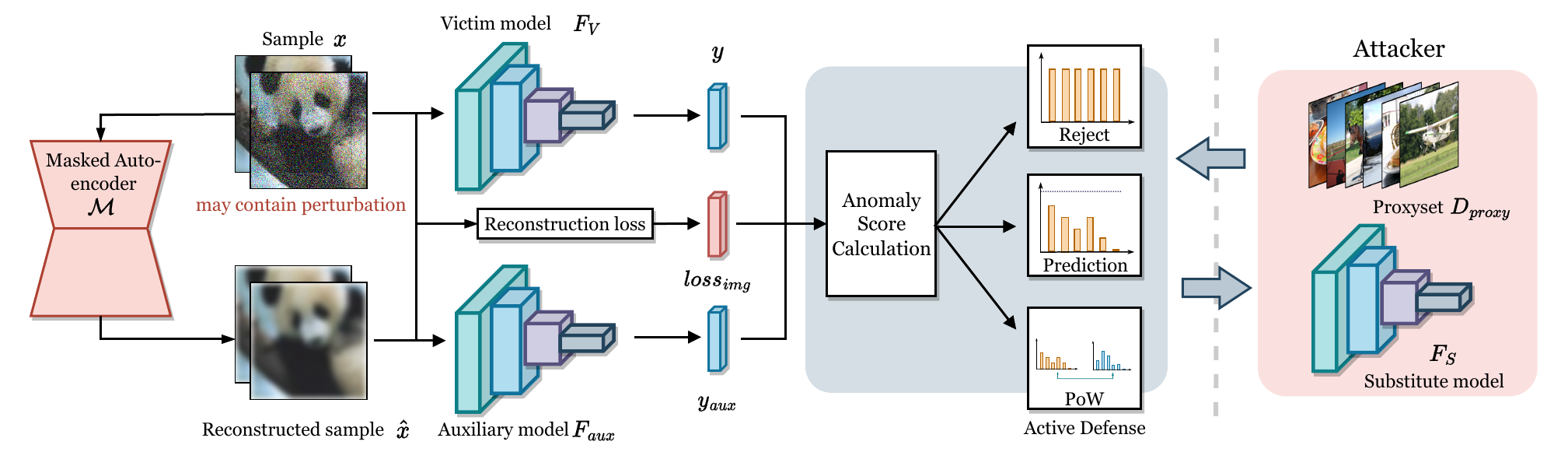}
    \caption{ The workflow of the proposed \Name. Whenever a sample is received, a fully trained masked auto-encoder first performs sample reconstruction. The reconstructed sample is then fed into an auxiliary model that outputs an auxiliary prediction. The overall anomaly score is calculated based on two samples and two predictions. After that, an appropriate response strategy is selected according to the anomaly score. The victim model remains frozen throughout the defense.}
    \label{fig:framework}
\end{figure*}

\section{Motivation}
The design of \Name is motivated by two observations. 

First, \textbf{sample reconstruction can better disclose anomaly than sample classification.}
Existing detection methods build DNN classifiers to detect suspicious samples. However, 
numerous studies have shown that even with the Outlier Exposure (OE) strategy, OOD detection can exhibit over-confidence in unseen OOD samples \cite{nguyen2015deep,li2023rethinking}. As depicted in Figure \ref{fig:compare} (left), the introduction of the OE strategy cannot effectively differentiate the majority of malicious queries from benign ones. 
Instead of directly building the detection classifier, our key insight is to reconstruct the query samples and identify the anomalies from the reconstruction process. This is based on the observation that the reconstruction loss is a better indicator of malicious samples than classifier-based confidence scores in the model extraction scenario. For the first time, we introduce the idea of autoencoder-based sample reconstruction for model extraction attack detection. 


Second, \textbf{an auxiliary model can better facilitate the OOD detection.}
We find that sample reconstruction is effective for sample-based stealing and synthetic-based stealing, but not perturbation-based stealing, wherein 
the attacker has a small number of original datasets, and the whole query set is perturbed on these seed samples. In other words, the reconstruction loss for perturbation-based queries is similar to the one for benign queries, as shown in Figure \ref{fig:compare-jbda} (left). 
%
To address it, we further introduce an auxiliary model to distinguish such perturbation-based queries. Specifically, we adopt a copy of the original victim model as the architecture of the auxiliary model but train it on a reconstructed version (using MAE) of the original dataset. In this way, for perturbation-based queries, we will obtain different predictions from the victim model and the auxiliary model. But for the other queries, the predictions tend to be consistent. Based on this, we can easily detect malicious queries by perturbation-based stealing, as shown in Figure \ref{fig:compare-jbda} (right).


\section{Methodology}

Based on the above analysis, we introduce \Name, a novel model extraction attack detection methodology. It uses a sample reconstruction strategy based on Masked Auto-encoder to disclose the malicious behaviors of query samples. It further integrates an Auxiliary Model to repair the model prediction and reinforce the detection results. Compared to prior works, \Name can minimize the requirements for the defender's capabilities: i.e., he does not need extra datasets or white-box access to the victim model. \Fref{fig:framework} shows the overview of \Name, which consists of three stages: (1) Sample Reconstruction via Masked Auto-encoder; (2) Attack Repairing via Auxiliary Model; (3) Anomaly Score Calculation. We will explain each stage in the following part.

\subsection{Sample Reconstruction via Masked Auto-encoder}


The first stage of our detection pipeline is reconstructing the query sample with an auto-encoder. 
Auto-encoders, as unsupervised neural network architectures, are primarily employed for dimensionality reduction and feature learning. Despite their effectiveness in many scenarios, traditional auto-encoders might not always capture the most salient features, especially when dealing with noisy datasets. 
To address this issue, we employ a \textit{masked autoencoder} (MAE) for sample reconstruction. A masked autoencoder introduces an additional masking operation during the encoding phase. Before feeding the input data to the encoder, a mask is applied, forcing the encoder to focus only on specific portions of the data. The mask essentially provides a form of inductive bias, directing the model to concentrate on potentially informative segments of the input.

In \Name, the masked autoencoder $\mathcal{M}$ consists of an encoder $f_\theta: \mathbb{R}^n \to \mathbb{R}^m$ and a decoder $g_\phi: \mathbb{R}^m \to \mathbb{R}^n$, parameterized as $\theta$ and $\phi$ respectively. Let $x \in D_V$ be a sample in the victim training set $D_V$, and $b\in \{0,1\}^n$ be the mask matrix that is sampled following a probability distribution. During the training process, each sample $x$ will be masked and then passed to the encoder to get the latent variable $z= f_\theta(b \odot x)$. Afterward, the latent variable will be passed to the decoder $g_\phi$ to get the reconstructed sample $\hat{x}=\mathcal{M}(x)=(g_\phi \circ f_\theta)(b \odot x)$. The objective function of $\mathcal{M}$ is to compute the following MSE loss between the original input $x$ and reconstructed sample $\hat{x}$:
\begin{equation}
    \label{eq:mae-loss}
    L_{MAE} = \frac{1}{|D_V|} \sum_{x \in D_V} \lVert x - \hat{x} \rVert^2 ,
\end{equation}
where the loss is minimized when the model is fully trained on the victim dataset. Afterward, $\mathcal{M}$ is used for score calculation and attack repair in the subsequent steps.

\subsection{Attack Repair via Auxiliary Model}

We denote the Auxiliary Model as $F_{aux}$, to repair and reinforce the anomaly detection. Given our original dataset $D_V$, the masked autoencoder $\mathcal{M}$ processes each sample to produce a reconstructed dataset $\hat{D_V}$, where every sample $\hat{x_i} \in \hat{D_V}$ corresponds to a sample $x_i \in D_V$. We then train $F_{aux}$ using the reconstructed samples $\hat{X}$, paired with their respective original labels $Y = \{y_1, y_2, ..., y_N\}$. The goal here is to ensure that the predictions of $F_{aux}$ on $\hat{D_V}$ align as closely as possible with the predictions of the original victim model on $D_V$. Thus, the objective function for training $F_{aux}$ can be defined as:
\begin{equation}
L_{aux}= \frac{1}{N} \sum_{i=1}^{N} \lVert F_V(x_i) - F_{aux}(\hat{x}_i) \rVert^2,   
\end{equation}
where $F_V$ denotes the victim model, and the objective is to minimize the squared difference between the predictions of $F_V$ and $F_{aux}$ across all samples. By achieving this, we aim to ensure that the Auxiliary Model is a faithful reflection of the victim model's behavior but operates in the transformed space of the reconstructed dataset $\hat{D_V}$.

\subsection{Anomaly Score Calculation}
The primary goal of \Name is to output an anomaly score that indicates the malicious level of the query sample. The score consists of two parts: (1) the reconstruction loss from the masked auto-encoder; (2) the deviation loss from the auxiliary model. The whole score could be calculated as follows:
\begin{equation}
    S(x) = \alpha \cdot \lVert x - \hat{x} \rVert^2 + (1-\alpha) \cdot \lVert F_V(x)-F_{aux}(\hat{x}) \rVert^2,
    \label{eq:score}
\end{equation}
where $\alpha$ is a hyperparameter to balance the two score items. We perform ablation studies in the next section to evaluate the impact of this hyperparameter. 

\subsubsection{Flexibility as an Add-on.}
After obtaining the anomaly score, the defender can choose subsequent active defense strategies to weaken the adversary's performance. Without loss of generality, we implement two common defense strategies: reject prediction and proof-of-work. The former rejects responding to queries with high anomaly scores, which can prevent the attacker from obtaining sensitive information. The latter requires users to complete a proof-of-work (PoW) before they can get the prediction. And the difficulty of the PoW problem is tied to the anomaly scores.


\section{Experiment}    
 
\subsection{Experimental Settings}

\paragraph{Datasets and Model Architectures.} We evaluate our scheme on two groups of datasets: 1) MNIST \cite{lecun1998gradient} and EMNIST-digits \cite{cohen2017emnist}; 2) CIFAR-10 and CIFAR-100 \cite{krizhevsky2009learning}. Specifically, the victim model is trained on MNIST and CIFAR-10, while EMNIST-digits and CIFAR-100 serve as datasets for the attacker. Since the comparison scheme requires auxiliary anomaly datasets, we use KMNIST \cite{clanuwat2018deep} for MNIST and Tiny ImageNet \cite{le2015tiny} for CIFAR-10. To evaluate the performance of the defense in extreme cases, the attacker adopts the same model structure for his substitute model as the victim model: Conv3 (three-layer CNN) \cite{lecun1989backpropagation} for MNIST and ResNet-18 for CIFAR-10. 
In all experiments, we use a MAE model based on the ViT-Tiny encoder \cite{dosovitskiy2020image}, trained for 500 epochs on the victim training set.

\paragraph{Attack Methods.} We use three different attack methods: (1) KnockoffNets (Knockoff): as a sampling-based attack, it uses reinforcement learning to choose samples from the proxy dataset. (2) Jacobian-Based Dataset Augmentation (JBDA): this method uses a Jacobian-based data augmentation algorithm to generate new samples from seed samples. We utilize a seed dataset comprising 200 images, with a perturbation step size $\lambda$ set to 0.1. (3) Data-Free Model Extraction (DFME): this method belongs to the synthesis-based category. The attacker does not need any proxy dataset, which will lead to a decrease in attack performance.

\paragraph{Baseline Methods.} We choose two SOTA defense solutions as discussed above: (1) Outlier Exposure (OE); (2) Ensemble-based defense (EDM). Both methods need the model defender to collect an auxiliary malicious dataset. Then the victim model is trained in an adversarial manner. In contrast, our method does not require any auxiliary datasets, which is a more practical assumption.

\paragraph{Metrics.} To quantitatively evaluate the performance, we adopt three metrics: Area Under the Receiver Operating Characteristic curve (AUROC), Area Under the Precision-Recall curve (AUPR), and the False Positive Rate at N\% true positive rate (FPRN). AUROC evaluates the overall performance, while AUPR focuses on precision and recall particularly in imbalanced datasets. Besides, FPRN can better measure the trade-off between sensitivity and specificity.

\subsection{Comparisons with the Baseline Methods}

\begin{table*}[ht]
\centering
\small
\resizebox{\linewidth}{!}{
\begin{tabular}{cc|c|ccc|ccc|ccc}
\Xhline{1.5pt}
\multirow{2}{*}{$B$} &
  \multirow{2}{*}{Method} &
  \multirow{2}{*}{$D_{OE}$} &
  \multicolumn{3}{c|}{JBDA} &
  \multicolumn{3}{c|}{Knockoff} &
  \multicolumn{3}{c}{DFME} \\
 &
   &
   &
  AUROC$\uparrow$ &
  AUPR$\uparrow$ &
  FPR95$\downarrow$ &
  AUROC$\uparrow$ &
  AUPR$\uparrow$ &
  FPR95$\downarrow$ &
  AUROC$\uparrow$ &
  AUPR$\uparrow$ &
  FPR95$\downarrow$ \\ \Xhline{1.5pt}
\multirow{3}{*}{1k} & OE  & KMNIST & 58.70 & 56.28 & 84.60 & 91.78 & 92.59 & 43.60 & 100.00          & 100.00          & 0.00          \\
                    & EDM & KMNIST & 66.06 & 59.73 & 79.60 & 86.01 & 86.70 & 64.20 & 100.00          & 100.00          & 0.00          \\
 &
  \Name &
  - &
  \textbf{93.30} &
  \textbf{92.95} &
  \textbf{27.40} &
  \textbf{99.37} &
  \textbf{99.47} &
  \textbf{1.30} &
  100.00 &
  100.00 &
  0.00 \\ \hline
\multirow{3}{*}{4k} & OE  & KMNIST & 60.05 & 56.79 & 83.50 & 91.70 & 92.24 & 42.90 & \textbf{100.00} & \textbf{100.00} & 0.03          \\
                    & EDM & KMNIST & 71.37 & 65.42 & 74.65 & 85.04 & 85.63 & 67.11 & 99.99           & 99.99           & \textbf{0.00} \\
 &
  \Name &
  - &
  \textbf{93.10} &
  \textbf{92.84} &
  \textbf{28.52} &
  \textbf{99.19} &
  \textbf{99.38} &
  \textbf{1.25} &
  99.91 &
  99.75 &
  0.15 \\
\Xhline{1.5pt}
\end{tabular}}
\caption{AUROC (\%), AUPR (\%), and FPR95 (\%) of different detection methods under three different attacks on MNIST. }
\label{tab:main-results-mnist}
\end{table*}

\begin{table*}[ht]
\centering
\small
\resizebox{\linewidth}{!}{
\begin{tabular}{cc|c|ccc|ccc|ccc}
\Xhline{1.5pt}
\multirow{2}{*}{$B$} &
  \multirow{2}{*}{Method} &
  \multirow{2}{*}{$D_{OE}$} &
  \multicolumn{3}{c|}{JBDA} &
  \multicolumn{3}{c|}{Knockoff} &
  \multicolumn{3}{c}{DFME} \\
 &
   &
   &
  AUROC$\uparrow$ &
  AUPR$\uparrow$ &
  FPR95$\downarrow$ &
  AUROC$\uparrow$ &
  AUPR$\uparrow$ &
  FPR95$\downarrow$ &
  AUROC$\uparrow$ &
  AUPR$\uparrow$ &
  FPR95$\downarrow$ \\ \Xhline{1.5pt}
\multirow{3}{*}{6k} &
  OE &
  {\fontsize{7.5}{9.5}\selectfont ImageNet-T} &
  74.53 &
  70.41 &
  64.37 &
  77.38 &
  72.74 &
  60.77 &
  74.02 &
  60.64 &
  44.30 \\
 &
  EDM &
  {\fontsize{7.5}{9.5}\selectfont ImageNet-T} &
  73.59 &
  69.50 &
  71.35 &
  71.72 &
  67.63 &
  72.52 &
  80.35 &
  73.55 &
  49.72 \\
 &
  \Name &
  - &
  \textbf{80.89} &
  \textbf{79.07} &
  \textbf{56.68} &
  \textbf{84.03} &
  \textbf{79.97} &
  \textbf{43.98} &
  \textbf{97.64} &
  \textbf{97.43} &
  \textbf{12.52} \\ \hline
\multirow{3}{*}{10k} &
  OE &
  {\fontsize{7.5}{9.5}\selectfont ImageNet-T} &
  73.68 &
  69.53 &
  63.14 &
  77.82 &
  72.96 &
  59.96 &
  76.17 &
  62.99 &
  43.28 \\
 &
  EDM &
  {\fontsize{7.5}{9.5}\selectfont ImageNet-T} &
  74.17 &
  69.89 &
  69.98 &
  71.27 &
  67.02 &
  73.03 &
  81.41 &
  75.31 &
  48.52 \\
 &
  \Name &
  - &
  \textbf{80.66} &
  \textbf{78.77} &
  \textbf{58.18} &
  \textbf{83.84} &
  \textbf{79.71} &
  \textbf{44.13} &
  \textbf{98.21} &
  \textbf{97.96} &
  \textbf{8.02} \\ \Xhline{1.5pt}
\end{tabular}}
\caption{AUROC (\%), AUPR (\%), and FPR95 (\%) of different detection methods under three different attacks on CIFAR-10. }
\label{tab:main-results-cifar}
\end{table*}

\subsubsection{Effectiveness.}
Across all attack methods and datasets, \Name demonstrates generally superior performance compared to the baseline methods OE and EDM, particularly in AUROC and AUPR, as shown in Tables \ref{tab:main-results-mnist} and \ref{tab:main-results-cifar}. In MNIST, \Name consistently outperforms the baseline methods, especially against JBDA and Knockoff attacks. For the result under the 1k query budget, \Name achieves an impressive AUROC score of 93.30\% and 99.37\% for JBDA and Knockoff respectively. Similarly, AUPR scores are significantly higher for \Name, especially compared to OE and EDM. Increasing the query budget to 4k does not lead to a substantial difference in the performance metrics for \Name, suggesting its robustness irrespective of the attack cost. 

The superiority of \Name is more pronounced on the CIFAR10 dataset, further emphasizing its strength. For instance, with a 6k query budget under the Knockoff attack, \Name achieves an AUROC of 92.53\% which is significantly higher than that of  OE (77.38\%) and EDM (71.72\%). For the DFME attack, the performance of \Name reaches a remarkable 100.00\% in AUROC and AUPR under the 6k query budget, a feat unmatched by the other defenses. Interestingly, for the 10k query budget, \Name still retains its lead, especially against the Knockoff and DFME attacks.

\Name manages to achieve the lowest FPR95 for most attacks and settings, which is essential for practical implementations. A low FPR ensures that normal queries are not mistakenly classified as malicious, which otherwise could interrupt the normal service or degrade the users' experience.

In summary, \Name demonstrates robustness and superiority against various attacks across both datasets, highlighting its potential as a reliable defense mechanism.

\subsubsection{Fidelity and Efficiency.}
We also evaluate the impact of different defense methods on the performance of the victim model, as shown in Table \ref{tab:classifier-acc}. It demonstrates the negative impact of the OE and EDM strategies on the victim model in accuracy degradation and long training time. Moreover, it is observed that the implementation of a structure-sharing policy between the victim model and the auxiliary model can decrease memory usage without harming victim model accuracy. In summary, \Name is more suitable for malicious query detection in the model-stealing scenario.


\begin{table}[t]
\centering
\resizebox{\linewidth}{!}{
\begin{tabular}{c|c|c}
\Xhline{1.5pt}
Method     & Model Accuracy (\%)$\uparrow$ & Training Time (s)$\downarrow$ \\ \hline
OE & 80.83                   & 1562.92                      \\
EDM         & 80.37          & 2014.80              \\ 
\Name         & \textbf{83.28}          & \textbf{170.44}              \\ \Xhline{1.5pt}
\end{tabular}}
\caption{Accuracy and training time of the victim model under OE, EDM, and \Name.}
\label{tab:classifier-acc}
\end{table}

\subsubsection{Flexibility as an Add-on.}
\begin{figure}[t]
    \centering
    \includegraphics[width=0.92\linewidth]{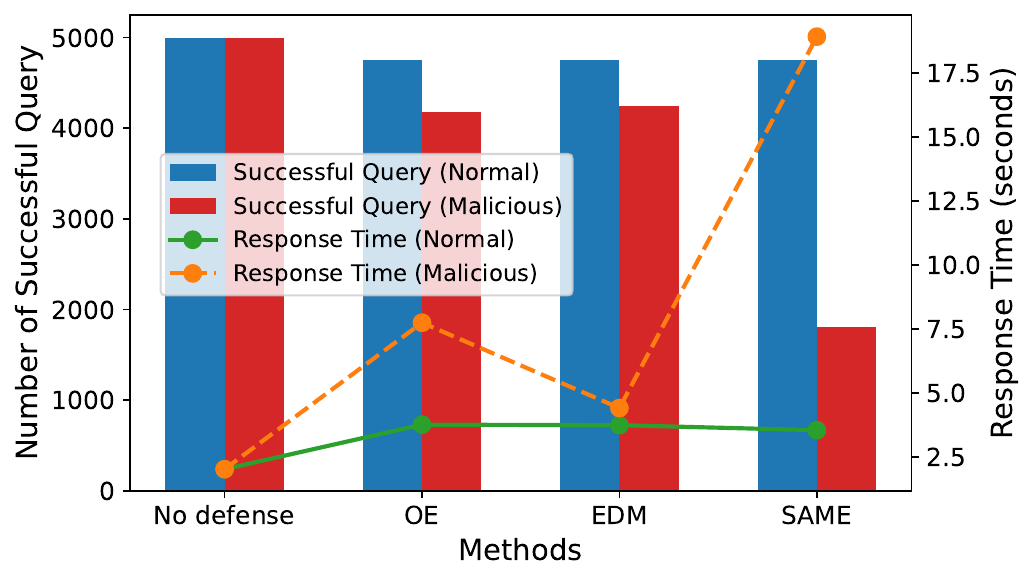}
    \caption{Comparison of flexibility as an add-on.}
    \label{fig:add-on}
\end{figure}

We test the performance of integrating \Name with two active defense methods: reject prediction and proof-of-work. Among them, reject prediction can be regarded as a special case of the perturbation-based defense. We use two metrics to evaluate the defense performance after splicing: the number of successful queries and response time, for both normal and malicious queries. As shown in Figure \ref{fig:add-on}, when using \Name as a detection plug-in, queries submitted by normal users are least negatively affected, while malicious queries are largely disturbed, on both metrics. This reflects the effectiveness and compatibility of \Name as a detection strategy.

\subsection{Ablation Study}



\begin{figure}[tb]
    \centering
    \includegraphics[width=\linewidth]{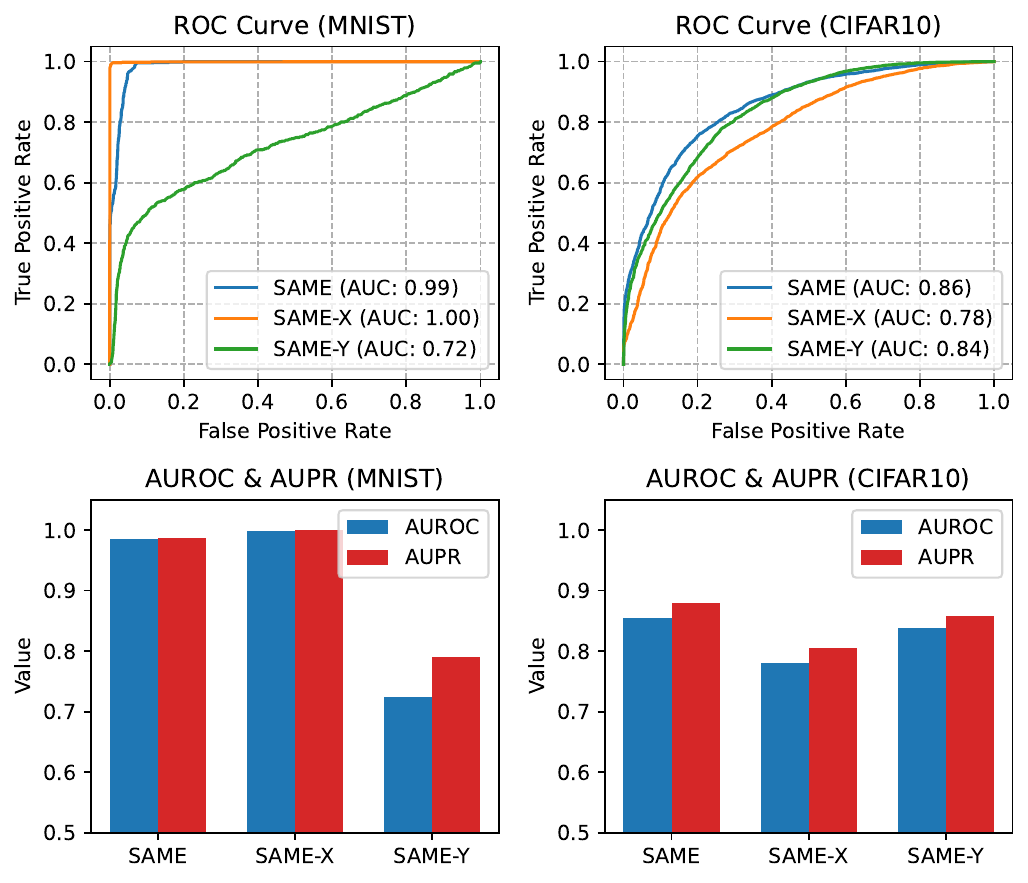}
    \caption{Detection performance of \Name and its variants on the MNIST and CIFAR-10 datasets.}
    \label{fig:ablation-roc}
\end{figure}

\paragraph{Effects of Different Components.} 
To demonstrate the effectiveness of each stage in \Name, we evaluate the performance of two variants of \Name: (1) \texttt{SAME-X}: only keeping the loss term based on sample reconstruction; (2) \texttt{SAME-Y}: only keeping the deviation loss item based on the Auxiliary Model. For fair comparisons, the masked autoencoders of the three \Name versions share the same weight, respectively, on MNIST and CIFAR-10. In addition, we adopt the same training configuration as in the previous section unless otherwise specified. As shown in Figure \ref{fig:ablation-roc}, on MNIST, \Name and \texttt{SAME-X} achieve competitive performance. We guess this is due to the simplicity of the dataset, which leads to the near-perfect reconstruction performance of MAE. On CIFAR-10, \Name also shows leading performance, while the other two variants perform close to the same.

\subsubsection{Effects of Malicious Ratio.}
\label{sec:ratio}
We further consider the scenario where the malicious and normal queries are unbalanced. We increase the ratio of malicious queries from 0.01 to 0.9 gradually, as shown in Table \ref{tab:ood-ratio}. For both MNIST and CIFAR-10, \Name's AUROC remains stable as the proportion of malicious samples increases. The AUPR value increases along with the malicious ratio for both datasets. This trend underscores the model's improved ability to identify malicious queries as the prevalence increases in the dataset.

\begin{table}[t]
\centering
\small
\resizebox{\linewidth}{!}{
\begin{tabular}{cc|cccc}
\Xhline{1.5pt}
Dataset                                             & Ratio & AUROC$\uparrow$ & AUPR$\uparrow$ & FPR95$\downarrow$ & FPR90$\downarrow$ \\ \Xhline{1.5pt}
\multirow{6}{*}{MNIST}    & 0.01  & 98.93           & 82.78          & 3.90              & 3.05              \\
                                                    & 0.05  & 99.07           & 95.42          & 3.82              & 3.12              \\
                                                    & 0.10  & 99.15           & 97.79          & 3.70              & 2.95              \\
                                                    & 0.30  & 99.23           & 99.30          & 3.65              & 2.80              \\
                                                    & 0.50  & 99.21           & 99.56          & 3.70              & 2.80              \\
                                                    & 0.90  & 99.21           & 99.71          & 3.70              & 2.80              \\ \hline
\multirow{6}{*}{CIFAR-10} & 0.01  & 86.91           & 50.91          & 61.20             & 46.62             \\
                                                    & 0.05  & 87.22           & 70.90          & 61.20             & 42.05             \\
                                                    & 0.10  & 87.64           & 79.53          & 60.45             & 40.58             \\
                                                    & 0.30  & 88.84           & 90.94          & 51.40             & 33.90             \\
                                                    & 0.50  & 89.39           & 94.23          & 47.20             & 31.45             \\
                                                    & 0.90  & 89.39           & 95.98          & 47.88             & 31.47             \\ \Xhline{1.5pt}
\end{tabular}}
\caption{The detection performance (\%) of \Name under different proportions of malicious samples.}
\label{tab:ood-ratio}
\end{table}

\begin{figure}[t]
    \centering
    \begin{subfigure}{\linewidth}
        \includegraphics[width=\linewidth]{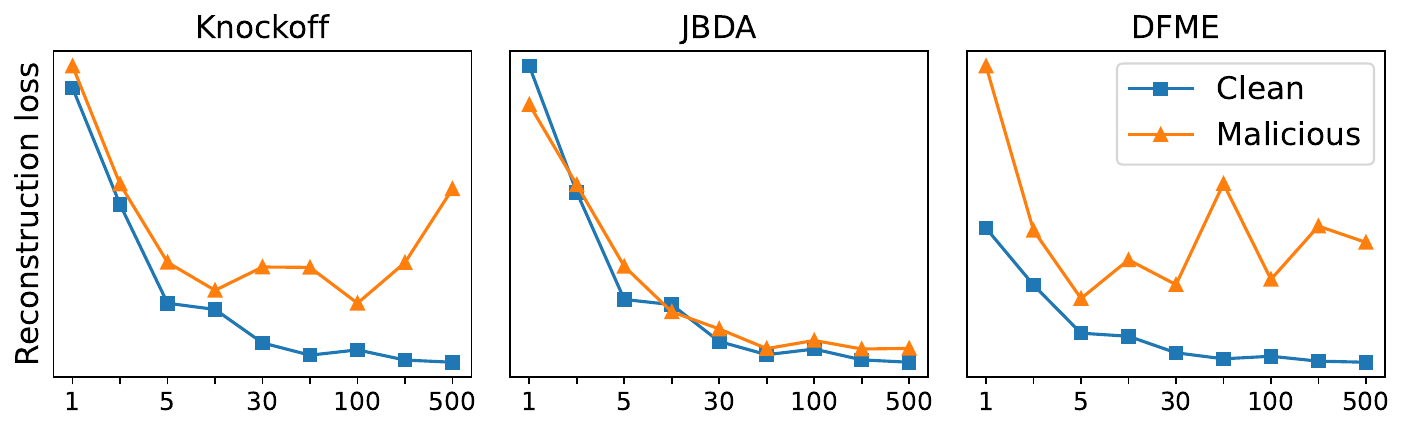}
    \end{subfigure}
    \begin{subfigure}{\linewidth}
        \includegraphics[width=\linewidth]{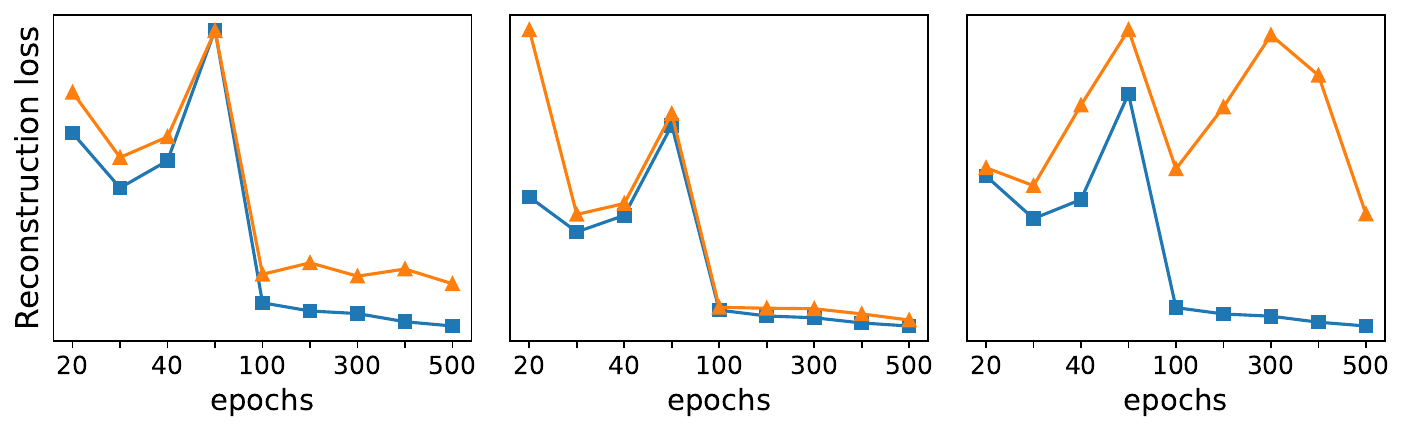}
    \end{subfigure}
    \caption{The reconstruction loss of the MAE for different categories of samples under different training epochs on MNIST (first row) and CIFAR-10 (second row) datasets.}
    \label{fig:epochs}
\end{figure}

\paragraph{Effects of MAE Training Epochs.} 
We evaluate the effect of the MAE training epochs on the reconstruction performance of clean and malicious samples, as shown in Figure \ref{fig:epochs}. MAE was trained for 500 epochs on MNIST and CIFAR10 datasets, with a 50-epoch warm-up. Post 100 epochs, the MAE's reconstruction ability stabilized, showing a distinct average loss for different sample types. With increasing epochs, the reconstruction loss gap for Knockoff and DFME attacks widened, attributed to MAE's improving reconstruction of the original dataset but not the attacker dataset (OOD). 
For JBDA attacks, involving minor data perturbations, MAE improved in reconstructing both clean and malicious samples, underscoring the importance of Auxiliary Model deviation loss in the anomaly score function (Equation \ref{eq:score}).



\begin{figure}[t]
    \centering
    \includegraphics[width=\linewidth]{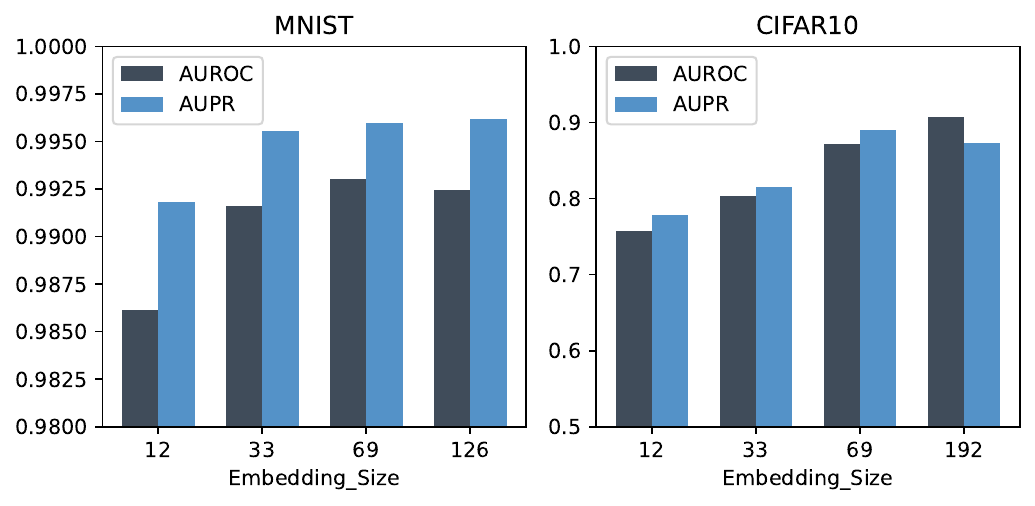}
    \caption{Detection performance of \Name under different MAE embedding sizes.}
    \label{fig:embedding_size}
\end{figure}

\paragraph{Effects of the MAE Embedding Size.} For MAE, its embedding size represents the dimensionality of the condensed representation obtained by the encoder. This size affects the information transfer fidelity between the encoder and decoder, and thus the model's reconstruction accuracy. 
In the experiment, deviation loss from the Auxiliary Model is excluded to avoid interference. Results in Figure \ref{fig:embedding_size} show that increased embedding size improves \Name's detection performance.


\section{Conclusion}
In this work, we propose a novel defense mechanism, \Name, to detect model extraction attacks. Compared to SOTA solutions, \Name does not require auxiliary datasets and demonstrates superior performance. Through comprehensive evaluations over common datasets, \Name displays high robustness against various extraction attacks under different query budgets.
Moreover, our ablation studies confirm the effectiveness of each stage of our proposed solution, emphasizing the significance of the embedded representation in the Masked Autoencoder and its impact on detection accuracy.
By integrating \Name with other active defenses, our end-to-end system exhibits improved defense capabilities. \Name enables maximum penalty for malicious queries while maintaining usability for normal users.
In the future, we aim to explore other variants of \Name and further optimize it for specific deployment scenarios. 
Additionally, studying its applicability across other types of machine learning models will also be a valuable avenue for future research.

\section{Acknowledgments}
We would like to thank the anonymous reviewers for their valuable comments and suggestions. 
This work is supported by 
the National Natural Science Foundation of China under Grant Nos. 61960206014 and 62121001, 
the National Research Foundation, Singapore, and the Cyber Security Agency under its National Cybersecurity R\&D Programme (NCRP25-P04-TAICeN), 
and Singapore Ministry of Education (MOE) AcRF Tier 2 MOE-T2EP20121-0006. 
Any opinions, findings and conclusions or recommendations expressed in this material are those of the authors and do not reflect the views of National Research Foundation, Singapore and Cyber Security Agency of Singapore.

\bibliography{aaai24}

\begin{thebibliography}{34}
\providecommand{\natexlab}[1]{#1}

\bibitem[{Barbalau et~al.(2020)Barbalau, Cosma, Ionescu, and Popescu}]{barbalau2020black}
Barbalau, A.; Cosma, A.; Ionescu, R.~T.; and Popescu, M. 2020.
\newblock Black-Box Ripper: Copying black-box models using generative evolutionary algorithms.
\newblock \emph{Advances in Neural Information Processing Systems}, 33: 20120--20129.

\bibitem[{Chandrasekaran et~al.(2020)Chandrasekaran, Chaudhuri, Giacomelli, Jha, and Yan}]{chandrasekaran2020exploring}
Chandrasekaran, V.; Chaudhuri, K.; Giacomelli, I.; Jha, S.; and Yan, S. 2020.
\newblock Exploring connections between active learning and model extraction.
\newblock In \emph{29th USENIX Security Symposium (USENIX Security 20)}, 1309--1326.

\bibitem[{Chen et~al.(2021)Chen, Guo, Zhang, Xie, and Liu}]{chen2021stealing}
Chen, K.; Guo, S.; Zhang, T.; Xie, X.; and Liu, Y. 2021.
\newblock Stealing deep reinforcement learning models for fun and profit.
\newblock In \emph{Proceedings of the 2021 ACM Asia Conference on Computer and Communications Security}, 307--319.

\bibitem[{Clanuwat et~al.(2018)Clanuwat, Bober-Irizar, Kitamoto, Lamb, Yamamoto, and Ha}]{clanuwat2018deep}
Clanuwat, T.; Bober-Irizar, M.; Kitamoto, A.; Lamb, A.; Yamamoto, K.; and Ha, D. 2018.
\newblock Deep learning for classical japanese literature.
\newblock \emph{arXiv preprint arXiv:1812.01718}.

\bibitem[{Cohen et~al.(2017)Cohen, Afshar, Tapson, and Van~Schaik}]{cohen2017emnist}
Cohen, G.; Afshar, S.; Tapson, J.; and Van~Schaik, A. 2017.
\newblock EMNIST: Extending MNIST to handwritten letters.
\newblock In \emph{2017 international joint conference on neural networks (IJCNN)}, 2921--2926. IEEE.

\bibitem[{Dosovitskiy et~al.(2020)Dosovitskiy, Beyer, Kolesnikov, Weissenborn, Zhai, Unterthiner, Dehghani, Minderer, Heigold, Gelly et~al.}]{dosovitskiy2020image}
Dosovitskiy, A.; Beyer, L.; Kolesnikov, A.; Weissenborn, D.; Zhai, X.; Unterthiner, T.; Dehghani, M.; Minderer, M.; Heigold, G.; Gelly, S.; et~al. 2020.
\newblock An image is worth 16x16 words: Transformers for image recognition at scale.
\newblock \emph{arXiv preprint arXiv:2010.11929}.

\bibitem[{Dziedzic et~al.(2021)Dziedzic, Kaleem, Lu, and Papernot}]{dziedzic2021increasing}
Dziedzic, A.; Kaleem, M.~A.; Lu, Y.~S.; and Papernot, N. 2021.
\newblock Increasing the Cost of Model Extraction with Calibrated Proof of Work.
\newblock In \emph{International Conference on Learning Representations}.

\bibitem[{He et~al.(2022)He, Chen, Xie, Li, Doll{\'a}r, and Girshick}]{he2022masked}
He, K.; Chen, X.; Xie, S.; Li, Y.; Doll{\'a}r, P.; and Girshick, R. 2022.
\newblock Masked autoencoders are scalable vision learners.
\newblock In \emph{Proceedings of the IEEE/CVF conference on computer vision and pattern recognition}, 16000--16009.

\bibitem[{Hendrycks, Mazeika, and Dietterich(2018)}]{hendrycks2018deep}
Hendrycks, D.; Mazeika, M.; and Dietterich, T. 2018.
\newblock Deep anomaly detection with outlier exposure.
\newblock \emph{arXiv preprint arXiv:1812.04606}.

\bibitem[{Jiang et~al.(2023)Jiang, Li, Xu, Zhang, and Lu}]{jiang2023comprehensive}
Jiang, W.; Li, H.; Xu, G.; Zhang, T.; and Lu, R. 2023.
\newblock A Comprehensive Defense Framework Against Model Extraction Attacks.
\newblock \emph{IEEE Transactions on Dependable and Secure Computing}.

\bibitem[{Juuti et~al.(2019)Juuti, Szyller, Marchal, and Asokan}]{juuti2019prada}
Juuti, M.; Szyller, S.; Marchal, S.; and Asokan, N. 2019.
\newblock PRADA: protecting against DNN model stealing attacks.
\newblock In \emph{2019 IEEE European Symposium on Security and Privacy (EuroS\&P)}, 512--527. IEEE.

\bibitem[{Kariyappa, Prakash, and Qureshi(2020)}]{kariyappa2020protecting}
Kariyappa, S.; Prakash, A.; and Qureshi, M.~K. 2020.
\newblock Protecting dnns from theft using an ensemble of diverse models.
\newblock In \emph{International Conference on Learning Representations}.

\bibitem[{Kariyappa, Prakash, and Qureshi(2021)}]{kariyappa2021maze}
Kariyappa, S.; Prakash, A.; and Qureshi, M.~K. 2021.
\newblock Maze: Data-free model stealing attack using zeroth-order gradient estimation.
\newblock In \emph{Proceedings of the IEEE/CVF Conference on Computer Vision and Pattern Recognition}, 13814--13823.

\bibitem[{Kariyappa and Qureshi(2020)}]{kariyappa2020defending}
Kariyappa, S.; and Qureshi, M.~K. 2020.
\newblock Defending against model stealing attacks with adaptive misinformation.
\newblock In \emph{Proceedings of the IEEE/CVF Conference on Computer Vision and Pattern Recognition}, 770--778.

\bibitem[{Krizhevsky, Hinton et~al.(2009)}]{krizhevsky2009learning}
Krizhevsky, A.; Hinton, G.; et~al. 2009.
\newblock Learning multiple layers of features from tiny images.

\bibitem[{Le and Yang(2015)}]{le2015tiny}
Le, Y.; and Yang, X. 2015.
\newblock Tiny imagenet visual recognition challenge.
\newblock \emph{CS 231N}, 7(7): 3.

\bibitem[{LeCun et~al.(1989)LeCun, Boser, Denker, Henderson, Howard, Hubbard, and Jackel}]{lecun1989backpropagation}
LeCun, Y.; Boser, B.; Denker, J.~S.; Henderson, D.; Howard, R.~E.; Hubbard, W.; and Jackel, L.~D. 1989.
\newblock Backpropagation applied to handwritten zip code recognition.
\newblock \emph{Neural computation}, 1(4): 541--551.

\bibitem[{LeCun et~al.(1998)LeCun, Bottou, Bengio, and Haffner}]{lecun1998gradient}
LeCun, Y.; Bottou, L.; Bengio, Y.; and Haffner, P. 1998.
\newblock Gradient-based learning applied to document recognition.
\newblock \emph{Proceedings of the IEEE}, 86(11): 2278--2324.

\bibitem[{Li et~al.(2022)Li, Xu, Guo, Qiu, Li, and Zhang}]{li2022extracting}
Li, G.; Xu, G.; Guo, S.; Qiu, H.; Li, J.; and Zhang, T. 2022.
\newblock Extracting Robust Models with Uncertain Examples.
\newblock In \emph{The Eleventh International Conference on Learning Representations}.

\bibitem[{Li et~al.(2023)Li, Chen, He, Yu, Liu, and Jia}]{li2023rethinking}
Li, J.; Chen, P.; He, Z.; Yu, S.; Liu, S.; and Jia, J. 2023.
\newblock Rethinking Out-of-distribution (OOD) Detection: Masked Image Modeling is All You Need.
\newblock In \emph{Proceedings of the IEEE/CVF Conference on Computer Vision and Pattern Recognition}, 11578--11589.

\bibitem[{Liu et~al.(2021)Liu, Ding, Shaham, Rahayu, Farokhi, and Lin}]{liu2021machine}
Liu, B.; Ding, M.; Shaham, S.; Rahayu, W.; Farokhi, F.; and Lin, Z. 2021.
\newblock When machine learning meets privacy: A survey and outlook.
\newblock \emph{ACM Computing Surveys (CSUR)}, 54(2): 1--36.

\bibitem[{Lou et~al.(2021)Lou, Guo, Li, Wu, and Zhang}]{lou2021naspy}
Lou, X.; Guo, S.; Li, J.; Wu, Y.; and Zhang, T. 2021.
\newblock NASPY: Automated extraction of automated machine learning models.
\newblock In \emph{International Conference on Learning Representations}.

\bibitem[{Nguyen, Yosinski, and Clune(2015)}]{nguyen2015deep}
Nguyen, A.; Yosinski, J.; and Clune, J. 2015.
\newblock Deep neural networks are easily fooled: High confidence predictions for unrecognizable images.
\newblock In \emph{Proceedings of the IEEE conference on computer vision and pattern recognition}, 427--436.

\bibitem[{Orekondy, Schiele, and Fritz(2019)}]{orekondy2019knockoff}
Orekondy, T.; Schiele, B.; and Fritz, M. 2019.
\newblock Knockoff nets: Stealing functionality of black-box models.
\newblock In \emph{Proceedings of the IEEE/CVF conference on computer vision and pattern recognition}, 4954--4963.

\bibitem[{Pal et~al.(2020)Pal, Gupta, Shukla, Kanade, Shevade, and Ganapathy}]{pal2020activethief}
Pal, S.; Gupta, Y.; Shukla, A.; Kanade, A.; Shevade, S.; and Ganapathy, V. 2020.
\newblock Activethief: Model extraction using active learning and unannotated public data.
\newblock In \emph{Proceedings of the AAAI Conference on Artificial Intelligence}, volume~34, 865--872.

\bibitem[{Papernot et~al.(2017)Papernot, McDaniel, Goodfellow, Jha, Celik, and Swami}]{papernot2017practical}
Papernot, N.; McDaniel, P.; Goodfellow, I.; Jha, S.; Celik, Z.~B.; and Swami, A. 2017.
\newblock Practical black-box attacks against machine learning.
\newblock In \emph{Proceedings of the 2017 ACM on Asia conference on computer and communications security}, 506--519.

\bibitem[{Shokri et~al.(2017)Shokri, Stronati, Song, and Shmatikov}]{shokri2017membership}
Shokri, R.; Stronati, M.; Song, C.; and Shmatikov, V. 2017.
\newblock Membership inference attacks against machine learning models.
\newblock In \emph{2017 IEEE symposium on security and privacy (SP)}, 3--18. IEEE.

\bibitem[{Tram{\`e}r et~al.(2016)Tram{\`e}r, Zhang, Juels, Reiter, and Ristenpart}]{tramer2016stealing}
Tram{\`e}r, F.; Zhang, F.; Juels, A.; Reiter, M.~K.; and Ristenpart, T. 2016.
\newblock Stealing machine learning models via prediction $\{$APIs$\}$.
\newblock In \emph{25th USENIX security symposium (USENIX Security 16)}, 601--618.

\bibitem[{Truong et~al.(2021)Truong, Maini, Walls, and Papernot}]{truong2021data}
Truong, J.-B.; Maini, P.; Walls, R.~J.; and Papernot, N. 2021.
\newblock Data-free model extraction.
\newblock In \emph{Proceedings of the IEEE/CVF conference on computer vision and pattern recognition}, 4771--4780.

\bibitem[{Wang et~al.(2021)Wang, Yin, Yao, Zhang, Fu, Ding, Li, Huang, and Xue}]{wang2021delving}
Wang, W.; Yin, B.; Yao, T.; Zhang, L.; Fu, Y.; Ding, S.; Li, J.; Huang, F.; and Xue, X. 2021.
\newblock Delving into data: Effectively substitute training for black-box attack.
\newblock In \emph{Proceedings of the IEEE/CVF Conference on Computer Vision and Pattern Recognition}, 4761--4770.

\bibitem[{Yang et~al.(2023)Yang, Hu, Cao, Xia, Huang, Liu, and Chen}]{yang2023protect}
Yang, Y.; Hu, M.; Cao, Y.; Xia, J.; Huang, Y.; Liu, Y.; and Chen, M. 2023.
\newblock Protect Federated Learning Against Backdoor Attacks via Data-Free Trigger Generation.
\newblock \emph{arXiv preprint arXiv:2308.11333}.

\bibitem[{Yao et~al.(2023)Yao, Li, Weng, Xue, Ren, and Qin}]{yao2023fdinet}
Yao, H.; Li, Z.; Weng, H.; Xue, F.; Ren, K.; and Qin, Z. 2023.
\newblock FDInet: Protecting against DNN Model Extraction via Feature Distortion Index.
\newblock \emph{arXiv preprint arXiv:2306.11338}.

\bibitem[{Yu et~al.(2020)Yu, Yang, Zhang, Tsai, Ho, and Jin}]{yu2020cloudleak}
Yu, H.; Yang, K.; Zhang, T.; Tsai, Y.-Y.; Ho, T.-Y.; and Jin, Y. 2020.
\newblock CloudLeak: Large-Scale Deep Learning Models Stealing Through Adversarial Examples.
\newblock In \emph{NDSS}.

\bibitem[{Zhao et~al.(2023)Zhao, Chen, Hao, Zhang, Xu, Li, and Zhang}]{zhao2023extracting}
Zhao, S.; Chen, K.; Hao, M.; Zhang, J.; Xu, G.; Li, H.; and Zhang, T. 2023.
\newblock Extracting Cloud-based Model with Prior Knowledge.
\newblock \emph{arXiv preprint arXiv:2306.04192}.

\end{thebibliography}

\end{document}